\documentclass[twocolumn,twoside,slac_two]{revtex4}
\usepackage{graphicx}
\usepackage{fancyhdr}
\input babarsym
\newcommand{\fL}{f_L}
\def\babar{{\em B}{\footnotesize\em A}{\em B}{\footnotesize\em AR}}
\newcommand{\etal}{{\it et al.}}
\newcommand{\thetaT}{\ensuremath{\theta_{\rm T}}}
\newcommand{\costhr}{\ensuremath{\cos\thetaT}}
\newcommand{\pvec}{{\bf p}}
\newcommand{\DE}{\ensuremath{\Delta E}}
\newcommand{\UfourS}{\mbox{$\Upsilon(4S)$}}

\newcommand{\calB}{\mbox{${\cal B}$}}
\newcommand{\bone}{\ensuremath{b_1}}
\newcommand{\Kst}{\ensuremath{K^*}}
\newcommand{\Kstp}{\ensuremath{\Kstarp}}
\newcommand{\Kstz}{\ensuremath{\Kstarz}}
\newcommand{\bonep}{\ensuremath{b_1^+}}
\newcommand{\bonem}{\ensuremath{b_1^-}}

\newcommand{\bonez}{\ensuremath{b_1^0}}
\newcommand{\rhop}{\ensuremath{\rho^+}}
\newcommand{\rhoz}{\ensuremath{\rho^0}}
\newcommand{\bmrhop}{\ensuremath{B^0 \rightarrow \fbmrhop}\xspace}

\newcommand{\rbmrhop}{\ensuremath{-1.8 \pm 0.5 \pm 1.0}}
\newcommand{\sbmrhop}{\ensuremath{-}}
\newcommand{\ulbmrhop}{\ensuremath{1.4}\xspace}

\newcommand{\rbprhoz}{\ensuremath{\msp 1.5\pm 1.5 \pm 2.2}}
\newcommand{\sbprhoz}{\ensuremath{0.4}}
\newcommand{\ulbprhoz}{\ensuremath{5.2}\xspace}

\newcommand{\rbzrhop}{\ensuremath{-3.0\pm 0.9 \pm 1.8}}
\newcommand{\sbzrhop}{\ensuremath{-}}
\newcommand{\ulbzrhop}{\ensuremath{3.3}\xspace}

\newcommand{\rbzrhoz}{\ensuremath{-1.1\pm 1.7^{+1.4}_{-0.9}}}
\newcommand{\sbzrhoz}{\ensuremath{-}}
\newcommand{\ulbzrhoz}{\ensuremath{3.4}\xspace}

\newcommand{\rbmKstpKppiz}{\ensuremath{\msp 1.8\pm 1.9 \pm 1.4}}
\newcommand{\sbmKstpKppiz}{\ensuremath{0.9}}

\newcommand{\rbmKstpKspip}{\ensuremath{\msp 3.2\pm 2.1^{+1.0}_{-1.5}}}
\newcommand{\sbmKstpKspip}{\ensuremath{1.5}}

\newcommand{\rbpKstz}{\ensuremath{\msp 2.9\pm 1.5 \pm 1.5}}
\newcommand{\sbpKstz}{\ensuremath{1.5}}
\newcommand{\ulbpKstz}{\ensuremath{5.9}\xspace}

\newcommand{\rbzKstpKppiz}{\ensuremath{-2.2\pm 3.0^{+5.0}_{-2.3}}}
\newcommand{\sbzKstpKppiz}{\ensuremath{-}}

\newcommand{\rbzKstpKspip}{\ensuremath{\msp 1.6\pm 2.5 \pm 3.3}}
\newcommand{\sbzKstpKspip}{\ensuremath{0.4}}

\newcommand{\rbzKstz}{\ensuremath{\msp 4.8\pm 1.9^{+1.5}_{-2.2}}}
\newcommand{\sbzKstz}{\ensuremath{2.0}}
\newcommand{\ulbzKstz}{\ensuremath{8.0}\xspace}

\newcommand{\rbmKstp}{\ensuremath{2.4^{+1.5}_{-1.3} \pm 1.0}}
\newcommand{\sbmKstp}{\ensuremath{1.7}}
\newcommand{\ulbmKstp}{\ensuremath{5.0}\xspace}

\newcommand{\rbzKstp}{\ensuremath{0.4^{+2.0+3.0}_{-1.5-2.6}}}
\newcommand{\sbzKstp}{\ensuremath{0.1}}
\newcommand{\ulbzKstp}{\ensuremath{6.7}\xspace}

\newcommand{\msp}{\ensuremath{\phantom{-}}}

\newcommand{\fbmrhop}{\ensuremath{\bonem\rhop}\xspace}

\newcommand{\fbprhoz}{\ensuremath{\bonep\rhoz}\xspace}

\newcommand{\fbzrhop}{\ensuremath{\bonez \rhop}\xspace}

\newcommand{\fbzrhoz}{\ensuremath{\bonez \rhoz}\xspace}

\newcommand{\fbpKstz}{\ensuremath{\bonep\Kstz}\xspace}

\newcommand{\fbmKstp}{\ensuremath{\bonem\Kstp}\xspace}

\newcommand{\fbmKstpKppiz}{\ensuremath{\bonem\Kstp_{\Kp\piz}}\xspace}

\newcommand{\fbmKstpKspip}{\ensuremath{\bonem\Kstp_{\KS\pip}}\xspace}

\newcommand{\fbzKstp}{\ensuremath{\bonez\Kstp}\xspace}

\newcommand{\fbzKstpKppiz}{\ensuremath{\bonez\Kstp_{\Kp\piz}}\xspace}

\newcommand{\fbzKstpKspip}{\ensuremath{\bonez\Kstp_{\KS\pip}}\xspace}

\newcommand{\fbzKstz}{\ensuremath{\bonez\Kstz}\xspace}

   \newcommand{\KstpKppiz}{\ensuremath{\Kstarp_{K^+\pi^0}}}
   \newcommand{\KstptoKppiz}{\ensuremath{\Kstarp\ra K^+\pi^0}}
   \newcommand{\KstpKspip}{\ensuremath{\Kstarp_{\KS\pi^+}}}
   \newcommand{\KstptoKspip}{\ensuremath{\Kstarp\ra \KS\pi^+}}
   \newcommand{\KstztoKppim}{\ensuremath{\Kstarz\ra K^+\pi^-}}
\newcommand{\acp}{\ensuremath{\calA_{ch}}}

\newcommand{\romegarhop}{\ensuremath{15.9\pm1.6\pm1.4}} % BF measurement
\newcommand{\somegarhop}{\ensuremath{9.8}}		% significance
\newcommand{\Aomegarhop}{\ensuremath{-0.20\pm0.09\pm0.02}}  % Ach
\newcommand{\fLomegarhop}{\ensuremath{0.90\pm0.05\pm0.03}}  % polarization

\newcommand{\romegarhoz}{\ensuremath{0.8\pm0.5\pm0.2}} % BF measurement
\newcommand{\ulomegarhoz}{\ensuremath{1.6}}	% 90% CL UL

\newcommand{\romegafz}{\ensuremath{1.0\pm0.3\pm0.1}} % BF measurement
\newcommand{\ulomegafz}{\ensuremath{1.5}}	% 90% CL UL

\newcommand{\romegaKstz}{\ensuremath{2.2\pm0.6\pm0.2}} % BF measurement
\newcommand{\somegaKstz}{\ensuremath{4.1}}		% significance
\newcommand{\AomegaKstz}{\ensuremath{0.45\pm0.25\pm0.02}}  % Ach
\newcommand{\fLomegaKstz}{\ensuremath{0.72\pm0.14\pm0.02}} % polarization

\newcommand{\romegaKpiSwavez}{\ensuremath{18.4\pm1.8\pm1.7}}

\newcommand{\somegaKpiSwavez}{\ensuremath{9.8}}	% significance

\newcommand{\AomegaKpiSwavez}{\ensuremath{-0.07\pm0.09\pm0.02}}

\newcommand{\romegaKtstz}{\ensuremath{10.1\pm2.0\pm1.1}}
\newcommand{\somegaKtstz}{\ensuremath{5.0}}		% significance
\newcommand{\AomegaKtstz}{\ensuremath{-0.37\pm0.17\pm0.02}}
\newcommand{\fLomegaKtstz}{\ensuremath{0.45\pm0.12\pm0.02}} % polarization

\newcommand{\romegaKstp}{\ensuremath{2.4\pm1.0\pm0.2}} % BF measurement
\newcommand{\somegaKstp}{\ensuremath{2.5}}		% significance
\providecommand{\ulomegaKstp}{\ensuremath{3.8}}

\newcommand{\AomegaKstp}{\ensuremath{0.29\pm0.35\pm0.02}}	% Ach
\newcommand{\fLomegaKstp}{\ensuremath{0.41\pm0.18\pm0.05}}      % polarization

\newcommand{\romegaKpiSwavep}{\ensuremath{27.5\pm3.0\pm2.6}} % BF measurement
\newcommand{\somegaKpiSwavep}{\ensuremath{9.2}}		% significance

\newcommand{\AomegaKpiSwavep}{\ensuremath{-0.10\pm0.09\pm0.02}}

\newcommand{\romegaKtstp}{\ensuremath{21.5\pm3.6\pm2.4}} % BF measurement
\newcommand{\somegaKtstp}{\ensuremath{6.1}}		% significance
\newcommand{\AomegaKtstp}{\ensuremath{0.14\pm0.15\pm0.02}}
\newcommand{\fLomegaKtstp}{\ensuremath{0.56\pm0.10\pm0.04}}      % polarization

\def\Kbar    {\kern 0.2em\overline{\kern -0.2em K}{}\xspace}

\def\Kz      {\ensuremath{K^0}\xspace}
\def\Kzb     {\ensuremath{\Kbar^0}\xspace}
\def\Kstar   {\ensuremath{K^{\ast}}\xspace}
\def\Kstarp  {\ensuremath{K^{\ast+}}\xspace}
\def\Kstarm  {\ensuremath{K^{\ast-}}\xspace}

\def\Kstarz  {\ensuremath{K^{\ast0}}\xspace}
\def\Kstarzb {\ensuremath{\Kbar^{\ast0}}\xspace}

\def\btoKstarzKstarzb {\ensuremath{\Bz \rightarrow \Kstarz \Kstarzb}}

\def\btoKstarpKstarz {\ensuremath{\Bp \rightarrow \Kstarzb\Kstarp}}

\def\btoKstarpKstarm {\ensuremath{\Bz \rightarrow \Kstarm\Kstarp}}

\newcommand{\kznsig}     {\mbox{$6.9^{+4.5}_{-3.5}$}}

\newcommand{\kzbfsyst}   {\mbox{$0.11$}}
\newcommand{\kzbf}       {\mbox{$0.85^{+0.61}_{-0.44}\pm \kzbfsyst $}}
\newcommand{\kzflstat}   {\mbox{$0.72^{+0.23}_{-0.36}$}}
\newcommand{\kzfl}       {\mbox{$\kzflstat\pm0.03$}}
\newcommand{\kzsubbf}    {\mbox{$15.37$}} % sub-branching fraction (%)
\newcommand{\kzsig}      {\mbox{$2.28$}} % significance
\newcommand{\kzbias}     {\mbox{$-0.12$}}

\newcommand{\kzeffavg}   {\mbox{$11.44\pm0.08$}}
\newcommand{\kpnsig}     {\mbox{$13.9^{+7.6}_{-6.4}$}}

\newcommand{\kpbfsyst}   {\mbox{$0.16$}}
\newcommand{\kpbf}       {\mbox{$1.80^{+1.01}_{-0.85}\pm\kpbfsyst$}}
\newcommand{\kpflstat}   {\mbox{$0.79^{+0.22}_{-0.36}$}}
\newcommand{\kpfl}       {\mbox{$\kpflstat\pm0.03$}}
\newcommand{\kpsubbf}    {\mbox{$22.22$}}

\newcommand{\kpsig}      {\mbox{$2.18$}} % significance
\newcommand{\kpbias}     {\mbox{$0.08$}}

\newcommand{\kpeffavg}   {\mbox{$7.40\pm0.08$}}

\newcommand{\kcbf}       {\mbox{$1.2\pm0.5\pm0.1$}}

\newcommand{\kcfl}       {\mbox{$0.75^{+0.16}_{-0.26}\pm0.03$}}
\newcommand{\kcup}       {\mbox{$2.0$}}
\newcommand{\kcsig}      {\mbox{$3.7$}}
\begin{document}

\title{Charmless hadronic B decays into Vector, Axial Vector and Tensor final states at BaBar}


\author{Paolo Gandini (On behalf of the \babar\ Collaboration)}
\affiliation{Universit\`a degli Studi and INFN Milano, via Celoria 16, I-20133 Milano, Italy}

\begin{abstract}
We present experimental measurements of branching fraction and longitudinal polarization fraction in charmless hadronic B
decays into vector, axial vector and tensor final states with the final dataset of \babar . Measurements of such
kind of decays are a powerful tool both to test the Standard Model and search possible sources of new physics.
\end{abstract}

\maketitle

\thispagestyle{fancy}

%%%%%%%%%%%%%%%%%%%%%%%%%%%%%%%%%%
\section{Introduction}
In this document we present a short review of the last experimental results at \babar\ concerning
charmless quasi two-body decays in final states containing particles with \textit{spin 1} or \textit{spin 2} and
different parities. This kind of decays has received considerable theoretical interest
in the last few years \cite{Cheng1, Cheng, Calderon} and this particular attention has
led to interesting experimental results at the current \textit{b-factories}. In fact,
the study of longitudinal polarization fraction $f_L$\ in charmless $B$ decays to vector 
vector ($VV$), vector axial-vector ($VA$) and axial-vector axial-vector ($AA$) mesons provides information
on the underlying helicity structure of the decay mechanism. Na\"{i}ve helicity conservation
arguments predict a dominant longitudinal polarization fraction $f_L \sim 1$\ for both
tree and penguin dominated decays and this pattern seems to be confirmed by
tree-dominated $B \ra \rho \rho$\ \cite{RhoRho} and $B^+ \ra \omega \rho^+$\ \cite{OmegaRho} decays.
Other penguin dominated decays, instead, show a different behavior: the measured value of
$\fL \sim 0.5$\ in $B \ra \phi K^*$\ decays \cite{PhiKst} is in contrast with na\"{i}ve Standard Model
(SM) calculations. Several solutions have been proposed such as the introduction of non-factorizable 
terms and penguin-annihilation amplitudes \cite{SMCal}, while other explanations invoke new physics \cite{NP}.
New modes have been investigated to shed more light on the problem.

\section{Helicity Amplitudes}
\begin{figure}[h]
\includegraphics[scale=0.3]{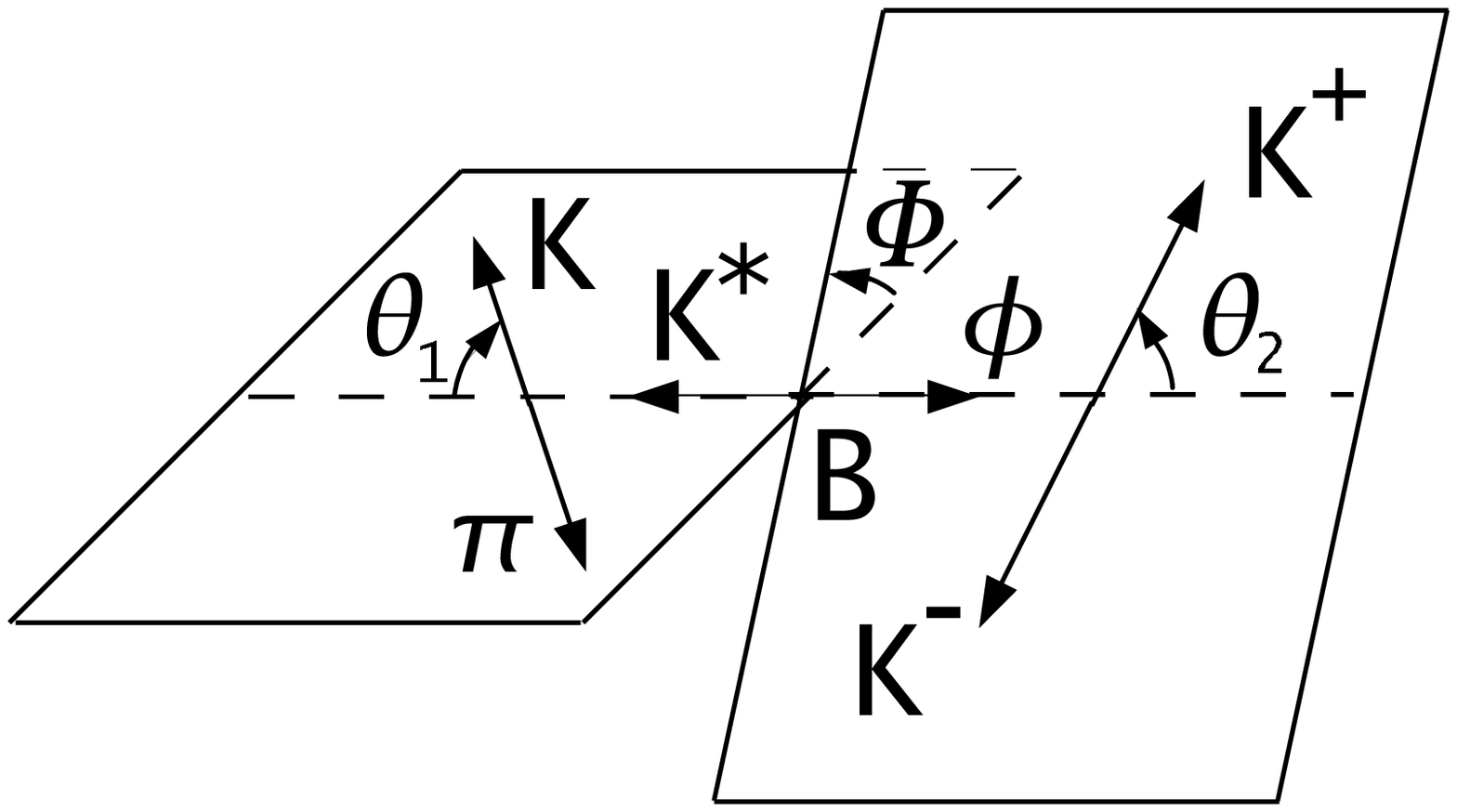}
\includegraphics[scale=0.3]{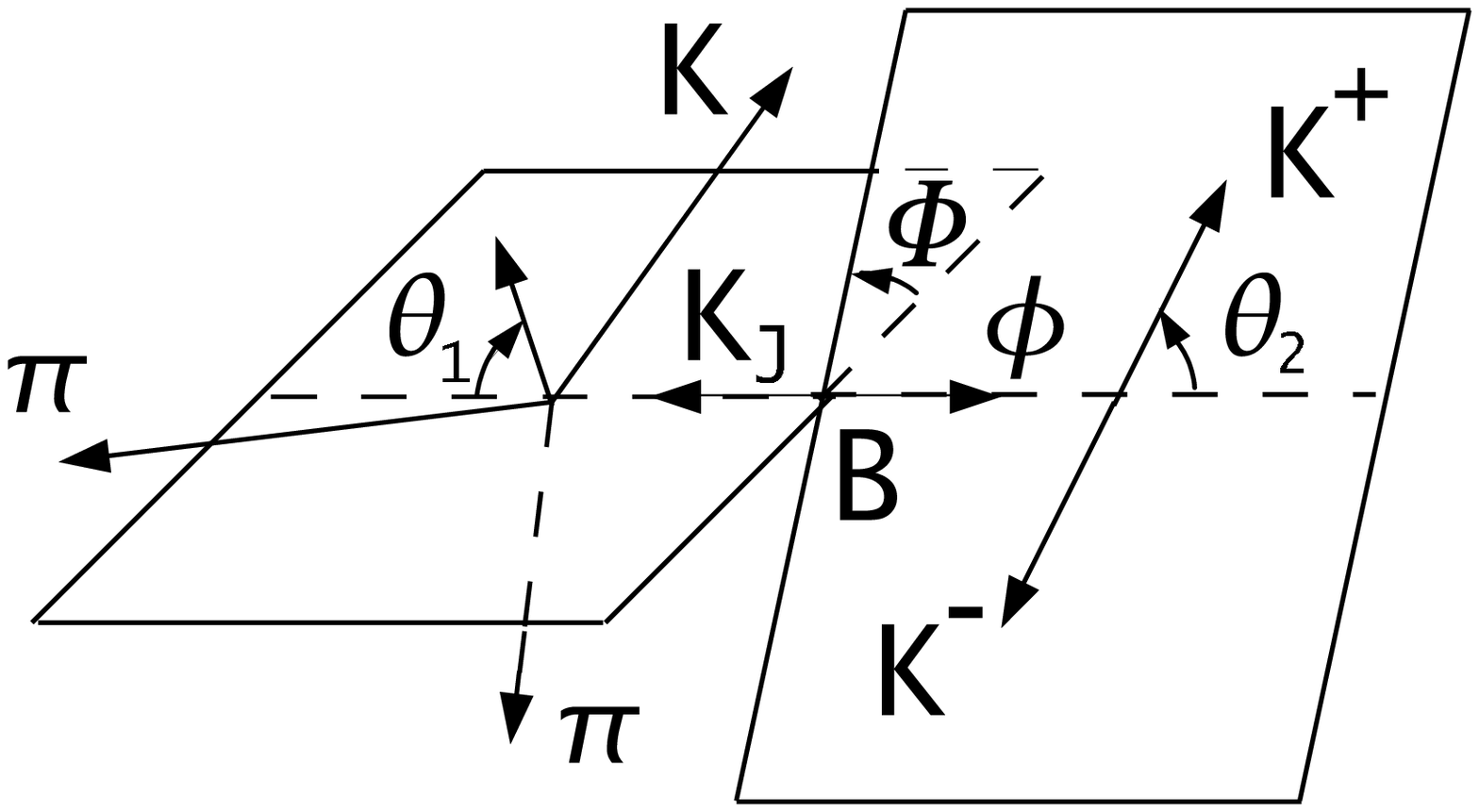}
\caption{\label{angoli}
Definition of the decay angles $\theta_1$\ and $\theta_2$\ given in the rest frames of the decaying parents for 2-body and 3-body decays.}
\end{figure}

The polarization fraction is extracted from data introducing some angular information
describing the decay process.
The easiest way to study angular distributions is in terms of helicity amplitudes:
total angular momentum must be conserved in $(Spin-0) \ra (Spin-1)+(Spin-1)$\ decays (similar
arguments can be used for $(Spin-0) \ra (Spin-1)+(Spin-2)$\ decays), so orbital $L$\ must be 0, 1 or 2
while ${\vec{J}}_{tot}$\ projection along the flight direction of the daughter mesons must be equal to 0.
This suggests that the helicities of both daughter mesons must be the same (1, 0 or -1); so the decay
process can be described by three different amplitudes $\Lambda_1$\ $\Lambda_0$\ and $\Lambda_{-1}$.
The polarization of the two intermediate mesons can be commonly measured introducing angular distributions
as shown in Fig.~\ref{angoli}.
For 2-body decays we define the angle $\theta_i$\ as the angle between the direction of the recoiling $B$\ and the
direction of one of the resonance daughters, while for 3-body decays we use the angle between the normal to the decay plane
with respect tho the other resonance daughter. Due to the limited number of expected signal events in charmless
hadronic decays we do not perform a full angular analysis of the decay and we integrate on the angle $\phi$\ between
the two planes of the decaying particles, leading to a non-trivial dependence on a single parameter
$$f_{L}=\frac{|\Lambda_0|^2}{\sum_{i=-1}^{1} |\Lambda_i|^2}$$
We find these angular distributions for our decays:
{\small

$$
\frac{\mbox{ d$\Gamma$}}{\mbox{ d$\cos\theta_1$d$\cos\theta2$}} \propto
$$
$$
\left\{
\begin{array}{ll}
B\rightarrow VV &         f_L\cos^2\theta_1\cos^2\theta_2 + \frac{1}{4}f_T\sin^2\theta_1\sin^2\theta_2\\
B\rightarrow AA &         f_L\sin^2\theta_1\sin^2\theta_2 + \frac{1}{4}(1+\cos^2\theta_1)(1+\cos^2\theta_2) \\
B\rightarrow VT &         f_L\frac{1}{3}\cos^2\theta_1(3\cos^2\theta_2-1)^2 + f_T\sin^2\theta_1\sin^2\theta_2\cos^2\theta_2\\
... \\
\end{array} \right.
$$}
where $f_L$\ and $f_T = 1-f_L$\ can be fitted from data.
%%%%%%%%%%%%%%%%%%%%%%%%%%%%%%%%%%%%%%%%%%%%%%%%%%%%%%%%%
%%%%%%%%%%%%%%%%%%%%%%%%%%%%%%%%%%%%%%%%%%%%%%%%%%%%%%%%%

\section{Analysis Techniques}
The results presented here are based on data collected with the \babar\ detector~\cite{BABARNIM}
at the PEP-II asymmetric-energy $e^+e^-$ collider~\cite{pep}, located at the SLAC National
Accelerator Laboratory. \BB\ pairs are recorded at the $\Upsilon (4S)$\ resonance at a center-of-mass
energy of $\sqrt{s}=10.58\ \gev$.

Background in our analyses arises primarily from random track combinations
in continuum events ($e^+e^- \ra\qqbar$, with $q = u, d, s, c$), where a pair of light quarks is produced.
We reduce this background by using optimized cuts on event shape variables, such as the
angle $\theta_T$\ between the thrust axis of the $B$\ candidate and the thrust axis of the rest of the event:
the distribution of $|\costhr|$ is sharply peaked near $1$ for combinations drawn from jet-like continuum
events and is nearly uniform for \BB\ events. Additional cuts are applied on the invariant masses
of the decaying particles, the $\chi^2$ probability of the $B$\ vertex fit, while we
impose particle identification requirements to ensure a good $K$/$\pi$/$p$\ separation.
A $B$ meson candidate is kinematically characterized by the energy-substituted
mass $\mes \equiv \sqrt{(s/2 + \pvec_0\cdot \pvec_B)^2/E_0^2 - \pvec_B^2}$ and
energy difference $\DE \equiv E_B^*-\sqrt{s}/2$, where the subscripts $0$ and
$B$\ refer to the initial \UfourS\ and the $B$ candidate in the laboratory frame, respectively,
and the asterisk denotes the \UfourS\ frame. $\mes$\ and $\DE$\ distributions are sharply
peaked for signal, while they are almost flat for \qqbar background.
Background can also arise from \BB\ events, which are taken into account with
detailed Monte Carlo simulations.

Results are obtained extracting the number of signal events from an unbinned extended maximum-likelihood (ML) fit
with input variables $\mes$, $\DE$, the invariant masses of the decaying particles, the helicity angles defined
above and the output of a Fisher discriminant (or a neural network) obtained combining different event shape variables.\\
The likelihood function is:
\begin{equation}
{\cal L}=  e^{-\left(\sum_{j=1}^{hyp} n_j\right)} \prod_{i=1}^N \left[\sum_{j=1}^{hyp} n_j  {\cal P}_j ({\bf x}_i)\right],
\end{equation}
where $N$ is the number of input events,  $n_j$ is the number of events for the hypothesis $j$\ (signal or background)
and     ${\cal P}_j ({\bf x}_i)$ is the corresponding probability density function (PDF), evaluated   
with the observables ${\bf x}_i$ of the $i$th event.

\section{Experimental Results}
%%%%%%%%%%%%%%%%%%%%%%%%%%%%%%%%%%%%%%%%%%%%%%%%%%%%%%%%%%%%%%%%%%%%%%%%%%%%%%%%%%%%%%%%%%%%%%%%%%%
\subsection{$B^0$ \ra  $a_1(1260)^+ a_1(1260)^-$}
We present the first measurements of the branching fraction and polarization 
in $B^0$ \ra  $a_1^+ a_1^-$ decays\footnote{$a_1$\ notation will be used to indicate the $a_1(1260)$\ meson.},
with $a_1^\pm$ decaying into three charged pions \cite{a1a1}.
For this mode, the only available experimental information is a
branching fraction upper limit (UL) of $2.8 \times 10^{-3}$\ at 90\%
confidence level (CL) measured by CLEO \cite{a1a1CLEO}, while there are no experimental measurements of $\fL$\
in $B \ra AA$\ decays.
Theoretical expectations for the branching fraction range from $37.4 \times 10^{-6}$\ \cite{Cheng} to $6.4 \times 10^{-6}$\ \cite{Calderon},
depending on the different approach used, while $\fL$\ is predicted to be about 0.64 \cite{Cheng}.
In this decay mode we reconstruct $B^0 \ra a_1^+ a_1^-$, with $a_1^\pm \ra \rho(770) \pi^\pm$\
and $\rho(770) \ra \pi^+ \pi^-$. We do not separate the P-wave $(\pi\pi)_{\rho}$
and the S-wave  $(\pi\pi)_{\sigma}$\ components in the $a_1 \rightarrow 3\pi$\ decay;
a systematic uncertainty is estimated due to the difference in the selection efficiencies.
The fit results, presented in Tab.~\ref{tab:resultsa1}, are based on an integrated
luminosity corresponding to 423 \invfb (equivalent to $(465\pm5)\times 10^6$ \BB\ pairs).

\begin{table}[!h]
\begin{center}
\begin{tabular}{lc}
\hline\hline
Signal yield & $545 \pm 118$ \\
Signal yield bias & $ +14$ \\
$\fL$\ bias & $ -0.06$\\
\hline
$S$ $(\sigma)$ & $5.0$ \\
$\calB\ (\times 10^{-6})$ & $47.3 \pm 10.5 \pm 6.3$ \\
$\fL$ & $0.31 \pm 0.22 \pm 0.10 $\\
\hline\hline 
\end{tabular}
\end{center}
\caption{Fitted signal yield and  yield bias (in events), bias on $\fL$, significance $S$
(including systematic uncertainties), measured branching
fraction \calB\ and fraction of longitudinal polarization $\fL$\ with statistical and systematic uncertainties.}
\label{tab:resultsa1}
\end{table}

\noindent
The significance is  the square root of the difference between the value of $-2\ln{\cal L}$ (with systematic 
uncertainties included) for zero signal and the value at its
minimum.
In this calculation we have taken into account the fact that the 
floating $\fL$\ parameter is not defined in the zero
signal hypothesis.
The measured branching fraction and longitudinal polarization are
in general agreement with QCD factorization expectations \cite{Cheng}.
%%%%%%%%%%%%%%%%%%%%%%%%%%%%%%%%%%%%%%%%%%%%%%%%%%%%%%%%%%%%%%%%%%%%%%%%%%%%%%%%%%%%%%%%%%%%%%%%%%%

\subsection{$B$ \ra  $b_1 V$\ \, with\ \,  $V = \rho, K^*$}
%%%%%%%%%%%%%%%%%%%%%%%%%%%%%%%%%%%%%%%%%%%%%%%%%%%%%%%%%%%%%%%%%%%%%%%%%%%%%%%%%%%%%%%%%%%%%%%%%%%
We search for all charge combinations of decays of a $B$ meson to a 
final state containing a $\bone$ meson and a $\rho$ or $\Kst(892)$ meson;
both neutral and charged $B$\ decays have been considered \cite{b1V}.
No previous experimental searches for these decays have been reported before.
Such $B$\ meson decays to charmless $AV$ final states are interesting to be studied experimentally
since they may be sensitive to penguin annihilation effects, which tend to enhance certain modes 
while suppressing others. Branching fractions for $AV$ modes are substantial in several cases,
as large as $33\times10^{-6}$ for the \bmrhop\ final state \cite{Cheng}, so they should be accessible
at \babar. Measurements described here are based on an integrated luminosity of
424 \invfb, equivalent to $(465\pm5)\times 10^6$ \BB\ pairs.\\
We reconstruct $B$-meson daughter candidates through the decays
$\bone\ra\omega\pi$ (we assume this branching fraction to be 100\%),
$\omega\ra\pip\pim\piz$, $\rhop\ra\pip\piz$, $\rho^0 \ra\pip\pim$,
$\Kstz\ra\Kp\pim$, and $\Kstp\ra\Kp\piz$ or $\KS\pip$.
Results are summarized in Tab.~\ref{tab:b1V}.
We do not observe any statistically significant signal for any of
the eight decay modes.

\begin{table}[h]
\begin{center}{\footnotesize
\begin{tabular}{l|c|c|c|c}
\hline\hline
Mode            & $Y$    &    $Y_0$   & $S$ $(\sigma)$  &\calB\ \& U.L $(10^{-6})$ \\
\hline
\fbmrhop        & $-33\pm10$ & $4\pm2$    & \sbmrhop  & \rbmrhop\ ($<$\ulbmrhop) \\
\fbzrhop        & $-18\pm5$ & $-4\pm2$        & \sbzrhop  & \rbzrhop\ ($<$\ulbzrhop) \\
\fbprhoz        & $37\pm25$ & $8\pm4$ & \sbprhoz  & \rbprhoz\ ($<$\ulbprhoz) \\
\fbzrhoz        & $-8\pm19$ & $5\pm3$    & \sbzrhoz  & \rbzrhoz\  ($<$\ulbzrhoz) \\
\hline
\fbmKstp         &        &       & \sbmKstp  & \rbmKstp\ ($<$\ulbmKstp) \\
~~\fbmKstpKppiz& $3\pm8$ & $-5\pm3$  & \sbmKstpKppiz  & \rbmKstpKppiz  \\
~~\fbmKstpKspip& $17\pm9$  & $4\pm2$  & \sbmKstpKspip\ & \rbmKstpKspip   \\
\fbzKstp        &        &       & \sbzKstp  & \rbzKstp\   ($<$\ulbzKstp) \\
~~\fbzKstpKppiz& $-8\pm7$   & $-3\pm2$   & \sbzKstpKppiz  & \rbzKstpKppiz  \\
~~\fbzKstpKspip& $3\pm4$ & $0\pm0$  & \sbzKstpKspip  & \rbzKstpKspip \\
\fbpKstz        & $55\pm21$   & $15\pm8$  & \sbpKstz  & \rbpKstz\  ($<$\ulbpKstz) \\
\fbzKstz        & $30\pm15$   & $-6\pm3$ & \sbzKstz  & \rbzKstz\  ($<$\ulbzKstz) \\
\hline\hline
\end{tabular}}
\end{center}
\caption{Signal yield $Y$\ (events) and its statistical uncertainty, bias $Y_0$\ (evts),
significance $S$ (including 
systematic uncertainties) and central value of the branching
fraction $\cal B$ with associated upper limit (U.L.) at 90\% C.L.}
\label{tab:b1V}
\end{table}

\noindent
These results are in good agreement with the small predictions from na\"{i}ve factorization calculations \cite{Calderon}, but
they are much smaller than the predictions from the more complete QCD factorization calculations \cite{Cheng}.

\subsection{$B \ra \omega V$\ \, with\ \, $V=\Kstar, \rho, f_0$}
We report measurements of $B$ decays to the 
final states $\omega\Kst$, $\omega\rho$, and $\omega f_0(980)$, where \Kstar\ 
includes the spin 0, 1, and 2 states, $\Kst_0(1430)$, \Kst(892), and $\Kst_2(1430)$, respectively~\cite{omegaV}.
The analyzed data sample corresponds to 465$\times10^6$ \BB\ pairs. We measure the branching fractions for nine of these decays,
five are observed for the first time; where relevant signal is found we also extract the direct
\CP-violating, time-integrated charge asymmetry and $\fL$.
$B$-daughter candidates are reconstructed through their decays $\rho^0\ra\pip\pim$, $f_0(980)\ra\pip\pim$, $\rho^+\ra\pip\piz$,
\KstztoKppim, \KstptoKppiz (\KstpKppiz), \KstptoKspip
(\KstpKspip), $\omega\ra\pip\pim\piz$, $\pi^0\ra\gaga$, and $K_s\ra\pip\pim$.
In Tab.~\ref{tab:OmegaV} we show for each decay mode the measured \calB, 
$\fL$, and \acp\ together with the quantities entering into these computations. 
For decays with \Kstarp\ we combine the results from the two \Kstar\ decay
channels, by adding their values of $-2\ln{\cal L}$.

%%%%%%%%%%%%%%%%%%%%%%%%%%%%%%%%%%%%%%%%%%%%%%%%%%%%%%%%%%%%%%%%%%%%%%%%%%%%%%%%%%%%%%%%%%%%%%%%%%%
\begin{table}[h]
\begin{center}{\footnotesize
\begin{tabular}{l|c|c|c}
\hline\hline
       & $\calB$ ($10^{-6}$) & $\calB$ \& UL ($10^{-6}$) & $S$ ($\sigma$)\\
\hline
$\omega K^*(892)^0$   & \romegaKstz &$-$ & \somegaKstz \\ 
$\omega K^*(892)^+$    & \romegaKstp &\ulomegaKstp &\somegaKstp\\ 

$\omega (K\pi)_0^{*0}$ &\romegaKpiSwavez & -- & \somegaKpiSwavez\\ 
$\omega (K\pi)_0^{*+}$ & \romegaKpiSwavep & -- & \somegaKpiSwavep\\ 
$\omega K_2(1430)^{*0}$ & \romegaKtstz & -- &\somegaKtstz \\ 
$\omega K_2(1430)^{*+}$ & \romegaKtstp& -- & \somegaKtstp \\ 
$\omega f_0$ & \romegafz   &\ulomegafz& 4.5 \\
$\omega\rho^0$ & \romegarhoz &\ulomegarhoz \\
$\omega\rho^+$ & \romegarhop &-- &\somegarhop \\
\hline
\end{tabular}

\begin{tabular}{l|c|c}
\hline
      & $A_{ch}$ & $f_L$\\
\hline
$\omega K^*(892)^0$   & +\AomegaKstz &\fLomegaKstz\\
$\omega K^*(892)^+$   &+\AomegaKstp &\fLomegaKstp \\ 
$\omega (K\pi)_0^{*0}$ &\AomegaKpiSwavez & --\\ 
$\omega (K\pi)_0^{*+}$ & \AomegaKpiSwavep & --\\ 
$\omega K_2(1430)^{*0}$& \AomegaKtstz&\fLomegaKtstz \\ 
$\omega K_2(1430)^{*+}$ & +\AomegaKtstp &\fLomegaKtstp \\
$\omega f_0$  & -- & -- \\
$\omega\rho^0$  & -- &0.8 fixed\\
$\omega\rho^+$ & \Aomegarhop  & \fLomegarhop \\
\hline\hline
\end{tabular}
}
\end{center}
\caption{Results for the modes presented in this section.
Up: central value of the branching fraction $\cal B$ with associated upper limit (U.L.) at 90\% C.L.
where available and significance $S$.
Down: charge asymmetry $A_{ch}$ and polarization fraction $f_L$.}
\label{tab:OmegaV}
\end{table}
%%%%%%%%%%%%%%%%%%%%%%%%%%%%%%%%%%%%%%%%%%%%%%%%%%%%%%%%%%%%%%%%%%%%%%%%%%%%%%%%%%%%%%%%%%%%%%%%%%%%

\subsection{\btoKstarpKstarz}
%%%%%%%%%%%%%%%%%%%%%%%%%%%%%%%%%%%%%%%%%%%%%%%%%%%%%%%%%%%%%%%%%%%%%%%%%%%%%%%%%%%%%%%%%%%%%%%%%%%
We present measurements of the branching fraction and longitudinal
polarization for the decay $B^+ \rightarrow \overline{K}^{*0} K^{*+}$,
with a sample of $467\pm5$ million $B\overline{B}$ pairs collected \cite{Kstar}.
The decay \btoKstarpKstarz\ occurs through both electroweak and
gluonic $b \rightarrow d$ penguin loops and its branching fraction
is expected to be of the same order as \btoKstarzKstarzb: theorethical
predictions based on QCD factorization range from $(0.5^{+0.2+0.4}_{-0.1-0.3}) \times 10^{-6}$~\cite{bib:Beneke06}
to $(0.6\pm0.1\pm 0.3) \times 10^{-6}$~\cite{Cheng}.
The \btoKstarzKstarzb\ branching fraction has been measured to be
$(1.28^{+0.35}_{-0.30} \pm 0.11) \times 10^{-6}$~\cite{bib:KstKst},
while an upper limit at the 90\% confidence level (C.L.) of
$2.0 \times 10^{-6}$ has been recently placed on the
\btoKstarpKstarm\ branching fraction~\cite{bib:KpKm}. The previous
experimental upper limit on the \btoKstarpKstarz\ branching fraction
at the 90\% C.L. is $71 (48) \times 10^{-6}$~\cite{bib:prevcleo},
assuming a fully longitudinally (transversely) polarized system.
The \btoKstarpKstarz\ candidates are reconstructed through the decays
of $\Kstarzb \to \Km\pip$ and $\Kstarp \to \KS\pip$ or 
$\Kstarp \to \Kp\piz$, with $\KS\to\pip\pim$ and
$\piz\to\gamma\gamma$.
The results of the ML fits are summarized in Tab.~\ref{tab:Kstar}.

\begin{table}[h]
\begin{center}{\footnotesize
\begin{tabular}{lcc}
\hline \hline
\noalign{\vskip1pt}
Final State  & \Km\pip \KS\pip & \Km\pip \Kp\piz \\
\hline
Yields (events):                   &   & \\
 Signal         & \kznsig & \kpnsig \\
 ML Fit Biases  & \kzbias  & \kpbias\ \\ \hline
Efficiencies and \calB: & & \\
 $\epsilon(\%)$ & \kzeffavg & \kpeffavg  \\
 $\prod\calB_{i} (\%)$ & \kzsubbf\ & \kpsubbf\ \\
\noalign{\vskip1pt}
 $f_L$\                   &    \kzfl &  \kpfl \\ 
\noalign{\vskip1pt}
 \calB\ ($\times 10^{-6}$) &\kzbf &\kpbf \\
\noalign{\vskip1pt}
 \calB\ Significance $S$ ($\sigma$) &  \kzsig  &  \kpsig \\ \hline
Combined Results: \\
 $f_L$\                   &  \multicolumn{2}{c}{\kcfl}\\ 
\noalign{\vskip1pt}
 \calB\ ($\times 10^{-6}$) & \multicolumn{2}{c}{\kcbf} \\
\noalign{\vskip1pt}
 \calB\ Significance $S$ ($\sigma$)      & \multicolumn{2}{c}{\kcsig} \\
 ${\cal B}_{\rm UL}$ ($\times 10^{-6}$) & \multicolumn{2}{c}{\kcup} \\
\hline
\hline
\end{tabular}}
\end{center}
\caption{Results of the fit: signal yield and ML Fit biases,
efficiencies and \calB\ for single and combined results.}
\label{tab:Kstar}
\end{table}

\noindent
We compute the branching fractions \calB\ by
dividing the bias-corrected yield by the number of \BB\ pairs, the
reconstruction efficiency $\epsilon$ given the fitted $f_L$, and the
secondary branching fractions, which we take to be $2/3$ for
$\calB(\Kstarzb \to \Km\pip)$ and $\calB(\Kstarp \to \Kz\pip)$, $1/3$
for $\calB(\Kstarp \to \Kp\piz)$, and $0.5 \times (69.20\pm0.05)\%$
for $\calB(\Kz\to\KS\to\pip\pim)$.
We see a significant excess of events, but no $5\sigma$\ observation is found;
all these measurements are compatible with theoretical predictions.

\begin{acknowledgments}
\noindent
I would like to thank Prof. Fernando Palombo and Dott. Vincenzo Lombardo for their support.
\end{acknowledgments}

\bigskip % extra skip inserted
%\begin{thebibliography}{9}   % Use for  1-9  references

\end{document}